\documentstyle[preprint,aps]{revtex}

\tightenlines

\begin{document}
\title{Exact Solution of Kemmer Equation for Coulomb Potential}
\author{S. G\"{o}nen$^{1}$, A. Havare$^{2\thanks{%
E-mail address: alihavare@mersin.edu.tr}}$, N. Unal$^{3}$}
\address{$^{1}$Dicle University, Department of Physics Education, Diyarbak\i r,\\
Turkey.\\
$^{2}$Mersin University, Department of Physics, Mersin, Turkey\\
$^{3}$Akdeniz University, Department of Physics, Antalya, Turkey}
\maketitle

\begin{abstract}
This article illustrates the bound states of Kemmer equation for spin-1
particles. The asymptotic, exact and Coulomb field solutions are obtained by
using action principle. In the conclusion the energy spectrum of spin-1
particles moving in a Coulomb potential compared with the energy spectrum of
spin-0 and spin-1/2 particles.
\end{abstract}

\section{Introduction}

\bigskip In the present article we analyzed the problem of massive, charged
particles of spin-1 in a Coulomb potential using action principle. The
quantum mechanics of charged, massive, spin-1 bosons in the presence of an
external field, was studied for many different situations using different \
techniques\cite{1,2,3,4}. These works especially included\ an investigation
of the solution of the equation in the presence of \ a magnetic field. These
techniques are far more difficult to employ when the anomalous magnetic
moment coupling terms are included into the equations of motion.

Relativistic spin-1 particle was at first described by Kemmer in 1939\cite{5}%
. Kemmer equation is a Dirac\ type equation, involves matrices obeying a
different scheme of commutation rules. It is reviwed because of the interest
in the quark-antiquark bound state problem. Corben and Schwinger were the
first to present a general theory for massive particles of spin-1 in an
external field \ and arbitrary magnetic moment \cite{6}. They generalized
the equations of Proca and studied the motion of such particles in an
external Coulomb field. Their work provides the basis for many other studies.

This article is the first that illustrates the exact solution of \ Kemmer
equation for massive spin-1 particle in the presence of a Coulomb potential.
The method used here is a generally accepted method to solve the
relativistic particle equations. The Lagrangian density of Kemmer
Hamiltonian was separated into radial and angular parts by using the spin
space rotation operators of $SO\left( 3\right) $ group in the action
principle \ and then the radial equations of motion are obtained by using
Euler-Lagrange equations of motion. Finally the asymptotic, exact and
Coulomb field solutions of these equations are discussed.

\section{Radial Lagrangian}

The Dirac-like relativistic Kemmer equation for spin-0 and spin-1 particle
of mass $m$ is 
\begin{equation}
\left( \beta ^{\mu }\pi _{\mu }-m\right) \Psi \left( x\right) =0  \label{1}
\end{equation}
where the $16\times 16$ Kemmer matrices $\beta ^{\mu }$ $\left( \mu
=0,1,2,3\right) $ satisfy the commutation relation 
\begin{equation}
\beta ^{\mu }\beta ^{\nu }\beta ^{\lambda }+\beta ^{\lambda }\beta ^{\nu
}\beta ^{\mu }=g^{\mu \nu }\beta ^{\lambda }+g^{\lambda \nu }\beta ^{\mu }
\label{2}
\end{equation}
and represented as

\begin{equation}
\beta ^{\mu }=\gamma ^{\mu }\otimes \text{I}+\text{I}\otimes \gamma ^{\mu }
\label{3}
\end{equation}
with $\gamma ^{\mu }$ usual Dirac matrices. The dynamical state $\Psi \left(
x\right) $ is a $16$-component\ and includes both spin-0 and spin-1
particles. $\pi _{\mu }$ is 4-vector electromagnetic momentum. After a basic
step Eq.(\ref{1}) can be written in the form

\begin{equation}
\left[ \left( \gamma ^{0}\otimes \text{I}+\text{I}\otimes \gamma ^{0}\right)
\pi _{0}-\left( \overrightarrow{\alpha }\otimes \gamma ^{0}+\gamma
^{0}\otimes \overrightarrow{\alpha }\right) \cdot \overrightarrow{\pi }\text{
}-m\gamma ^{0}\otimes \gamma ^{0}\right] \Psi \left( x\right) =0\text{ .}
\label{4}
\end{equation}
where $\overrightarrow{\alpha }$ are $4\times 4$ matrices and given in terms
of $2\times 2$ Pauli matrices as \qquad

\begin{equation}
\overrightarrow{\alpha }=\gamma ^{0}\overrightarrow{\gamma }=\left( 
\begin{array}{cc}
0 & \overrightarrow{\sigma } \\ 
\overrightarrow{\sigma } & 0
\end{array}
\right) .  \label{5}
\end{equation}

The Lagrangian density of this Hamiltonian is 
\begin{equation}
L\left( x\right) =\Psi ^{\dagger }\left[ \left( \gamma ^{0}\otimes \text{I}+%
\text{I}\otimes \gamma ^{0}\right) \pi _{0}-\left( \overrightarrow{\alpha }%
\otimes \gamma ^{0}+\gamma ^{0}\otimes \overrightarrow{\alpha }\right) \cdot 
\overrightarrow{\pi }\text{ }-m\gamma ^{0}\otimes \gamma ^{0}\right] \Psi
\left( x\right) \text{ .}  \label{6}
\end{equation}

This form must be separated into radial and angular parts to obtain the
radial equations. The approach of separation used here is a technique
employed by Barut and \"{U}nal \cite{7}, which introduces a rotation
aligning the z-axis of the coordinate system with the radial direction $%
\widehat{r}$. In the identical-two-fermion system spin space the rotation
operator is 
\begin{equation}
S_{R}=e^{\frac{i}{2}\left( \varphi \sigma ^{1}-\theta \sigma ^{3}\right) }
\label{7}
\end{equation}
where $\sigma ^{1}$ is the first component and $\sigma ^{3}$ is the third
component of Pauli matrices.

This operator preserves the wavefunction resulting 
\begin{equation}
\Psi \left( \overrightarrow{r}\right) \rightarrow \Psi ^{%
{\acute{}}%
}\left( \overrightarrow{r}\right) =S_{R}\left( 
\begin{array}{c}
-a\left( r,\theta ,\varphi \right) \\ 
ib\left( r,\theta ,\varphi \right) \\ 
ic\left( r,\theta ,\varphi \right) \\ 
d\left( r,\theta ,\varphi \right)
\end{array}
\right)  \label{8}
\end{equation}
where a,b,c,d are 4-component spinors. Hence, Eq.$\left( \ref{6}\right) $
takes the following form: 
\[
L\left( \overrightarrow{r}\right) =\left( -a^{\dagger },-ib^{\dagger
},-ic^{\dagger },d^{\dagger }\right) S_{R}^{-1}[\left( \gamma ^{0}\otimes 
\text{I}+\text{I}\otimes \gamma ^{0}\right) \pi _{0}- 
\]
\begin{equation}
\left( \overrightarrow{\alpha }\otimes \gamma ^{0}+\gamma ^{0}\otimes 
\overrightarrow{\alpha }\right) \cdot \overrightarrow{\pi }\text{ }-m\gamma
^{0}\otimes \gamma ^{0}]S_{R}\left( 
\begin{array}{c}
-a \\ 
ib \\ 
ic \\ 
d
\end{array}
\right) \text{ \ \ .}  \label{9}
\end{equation}
The 3-space rotation has the following effect in the spin space:

\begin{equation}
S_{R}^{-1}\left( 
\begin{array}{c}
\alpha ^{\theta } \\ 
\alpha ^{\varphi } \\ 
\alpha ^{r}
\end{array}
\right) S_{R}=\left( 
\begin{array}{c}
\alpha ^{1} \\ 
\alpha ^{2} \\ 
\alpha ^{3}
\end{array}
\right) \text{ .}  \label{10}
\end{equation}
Therefore, in the square paranthesis of the Lagrangian $\alpha _{r},$ $%
\alpha _{\theta },$ $\alpha _{\varphi }$ are rotated into $\alpha
_{1},\alpha _{2},\alpha _{3}$ respectively. Then Lagrangian density can be
rewritten as

\[
L\left( \overrightarrow{r}\right) =\left( -a^{\dagger },-ib^{\dagger
},-ic^{\dagger },d^{\dagger }\right) \{\left( \gamma ^{0}\otimes \text{I}+%
\text{I}\otimes \gamma ^{0}\right) \pi _{0}+i[(\gamma ^{0}\otimes \alpha
^{3}+\alpha ^{3}\otimes \gamma ^{0})\partial _{r} 
\]
\[
+\frac{1}{r}(\gamma ^{0}\otimes \alpha ^{+}\left( \sigma ^{-}-\partial
_{-}\right) +\gamma ^{0}\otimes \alpha ^{-}\left( \partial _{+}-\sigma
^{+}\right) +\alpha ^{+}\left( \sigma ^{-}-\partial _{-}\right) \otimes
\gamma ^{0} 
\]
\begin{equation}
+\alpha ^{-}\left( \partial _{+}-\sigma ^{+}\right) \otimes \gamma
^{0})]-m\gamma ^{0}\otimes \gamma ^{0}\}\left( 
\begin{array}{c}
-a \\ 
ib \\ 
ic \\ 
d
\end{array}
\right) \text{ \ .}  \label{11}
\end{equation}
The $\partial _{+}$ and $\partial _{-}$ operators in Lagrangian are helicity
lowering and raising operators, respectively and given by

\begin{eqnarray}
\partial _{+} &=&\partial _{\theta }+\frac{i}{\sin \theta }\partial
_{\varphi }+\frac{1}{2}\sigma ^{3}\cot \theta  \nonumber \\
\partial _{-} &=&-\partial _{\theta }+\frac{i}{\sin \theta }\partial
_{\varphi }+\frac{1}{2}\sigma ^{3}\cot \theta  \label{12}
\end{eqnarray}
To evaluate the matrix products in Lagrangian we first define new $4\times 4$
quantities in bloch form:

\begin{eqnarray}
\overrightarrow{\sigma }_{1} &=&\overrightarrow{\sigma }\otimes \text{I} 
\nonumber \\
\overrightarrow{\sigma }_{2} &=&\text{I}\otimes \overrightarrow{\sigma }
\label{13}
\end{eqnarray}
where I is a $2\times 2$ unit matrix, and

\begin{eqnarray}
\sigma ^{+} &=&\frac{1}{2}\left( \sigma ^{1}+i\sigma ^{2}\right)  \nonumber
\\
\sigma ^{-} &=&\frac{1}{2}\left( \sigma ^{1}-i\sigma ^{2}\right)  \label{14}
\end{eqnarray}
Then we can define $16\times 16$ matrices as follows:

\begin{eqnarray}
\overrightarrow{\alpha }_{\left( 1\right) } &=&\overrightarrow{\alpha }%
\otimes \text{I}  \nonumber \\
\overrightarrow{\alpha }_{\left( 2\right) } &=&\text{I}\otimes 
\overrightarrow{\alpha }  \nonumber \\
\gamma _{\left( 1\right) }^{\mu } &=&\gamma ^{\mu }\otimes \text{I} \\
\gamma _{\left( 2\right) }^{\mu } &=&\text{I}\otimes \gamma ^{\mu } 
\nonumber
\end{eqnarray}
where I is now $4\times 4$ unit matrix. In terms of these we can write the
following matrices 
\begin{eqnarray}
\alpha _{\left( 1\right) }^{\pm } &=&\frac{1}{2}\left( \alpha _{\left(
1\right) }^{1}\pm i\alpha _{\left( 1\right) }^{2}\right)  \nonumber \\
\alpha _{\left( 2\right) }^{\pm } &=&\frac{1}{2}\left( \alpha _{\left(
2\right) }^{1}\pm i\alpha _{\left( 2\right) }^{2}\right)  \label{16}
\end{eqnarray}
Straightforward algebra allows us to rewrite the Lagrangian in the form: 
\[
L\left( \overrightarrow{r}\right) =\{2\pi _{0}a^{\dagger }a-2\pi
_{0}d^{\dagger }d+[a^{\dagger }\sigma _{2}^{3}b+a^{\dagger }\sigma
_{1}^{3}c-b^{\dagger }\sigma _{2}^{3}c-b^{\dagger }\sigma
_{2}^{3}a-b^{\dagger }\sigma _{1}^{3}d 
\]
\[
-c^{+}\sigma _{1}^{3}a-c^{+}\sigma _{2}^{3}d+d^{+}\sigma
_{1}^{3}b+d^{+}\sigma _{2}^{3}c](\partial _{r}+\frac{1}{r})+\frac{i}{2r}%
[a^{+}\left( \overrightarrow{\sigma }_{1}\times \overrightarrow{\sigma }%
_{2}\right) ^{3}b- 
\]
\[
a^{+}\left( \overrightarrow{\sigma }_{1}\times \overrightarrow{\sigma }%
_{2}\right) ^{3}c-b^{\dagger }\left( \overrightarrow{\sigma }_{1}\times 
\overrightarrow{\sigma }_{2}\right) ^{3}a+b^{+}\left( \overrightarrow{\sigma 
}_{1}\times \overrightarrow{\sigma }_{2}\right) ^{3}d+c^{+}\left( 
\overrightarrow{\sigma }_{1}\times \overrightarrow{\sigma }_{2}\right) ^{3}a 
\]
\[
-c^{+}\left( \overrightarrow{\sigma }_{1}\times \overrightarrow{\sigma }%
_{2}\right) ^{3}d-d^{+}\left( \overrightarrow{\sigma }_{1}\times 
\overrightarrow{\sigma }_{2}\right) ^{3}b+d^{+}\left( \overrightarrow{\sigma 
}_{1}\times \overrightarrow{\sigma }_{2}\right) ^{3}c]-\frac{1}{r}%
[-a^{\dagger }\partial _{+}\sigma _{2}^{-}b- 
\]
\[
-a^{\dagger }\partial _{+}\sigma _{1}^{-}c+b^{\dagger }\partial _{+}\sigma
_{2}^{-}a+b^{\dagger }\partial _{+}\sigma _{1}^{-}d+c^{\dagger }\partial
_{+}\sigma _{1}^{-}a+c^{\dagger }\partial _{+}\sigma _{2}^{-}d-d^{\dagger
}\partial _{+}\sigma _{1}^{-}b-d^{\dagger }\partial _{+}\sigma _{2}^{-}c] 
\]
\[
+\frac{1}{r}[-a^{\dagger }\partial _{-}\sigma _{2}^{+}b-a^{\dagger }\partial
_{-}\sigma _{1}^{+}c+b^{\dagger }\partial _{-}\sigma _{2}^{+}a+b^{\dagger
}\partial _{-}\sigma _{1}^{+}d+c^{\dagger }\partial _{-}\sigma
_{1}^{+}a+c^{\dagger }\partial _{-}\sigma _{2}^{+}d 
\]
\begin{equation}
-d^{\dagger }\partial _{-}\sigma _{1}^{+}b-d^{\dagger }\partial _{-}\sigma
_{2}^{+}c]-ma^{\dagger }a+mb^{\dagger }b+mc^{\dagger }c-md^{\dagger }d\}%
\text{ \ .}  \label{17}
\end{equation}
where everything is 16-dimensional. One can verify that the square of total
angular momentum in spherical coordinates 
\begin{equation}
\text{J}^{2}=\{\partial _{\theta }^{2}+\cot \theta \partial _{\theta }+\frac{%
1}{\sin ^{2}\theta }\partial _{\varphi }^{2}+\frac{\cot ^{2}\theta }{4}\}+%
\text{J}_{r}^{2},
\end{equation}
can be written in terms of $\partial _{+}$ and $\partial _{-}$ operators as 
\begin{equation}
\text{J}^{2}=\frac{1}{2}[\partial _{+}^{\dagger }\partial _{+}+\partial
_{-}^{\dagger }\partial _{-}]+\text{J}_{r}^{2}
\end{equation}
This allows us to write the 4-component a,b,c,d bispinors in terms of
angular functions: 
\begin{equation}
a=\frac{1}{4\pi }\sum_{j,m}\left( 2j+1\right) \left( 
\begin{array}{c}
a_{+}\left( j,m;r\right) D_{+1,m}^{j}\left( \theta ,\varphi \right) \\ 
a_{0}\left( j,m;r\right) D_{0,m}^{j}\left( \theta ,\varphi \right) \\ 
a_{\tilde{o}}\left( j,m;r\right) D_{\tilde{o},m}^{j}\left( \theta ,\varphi
\right) \\ 
a_{-}\left( j,m;r\right) D_{-1,m}^{j}\left( \theta ,\varphi \right)
\end{array}
\right)  \label{20}
\end{equation}
\begin{equation}
b=\frac{1}{4\pi }\sum_{j,m}\left( 2j+1\right) \left( 
\begin{array}{c}
b_{+}\left( j,m;r\right) D_{+1,m}^{j}\left( \theta ,\varphi \right) \\ 
b_{0}\left( j,m;r\right) D_{0,m}^{j}\left( \theta ,\varphi \right) \\ 
b_{\tilde{o}}\left( j,m;r\right) D_{\tilde{o},m}^{j}\left( \theta ,\varphi
\right) \\ 
b_{-}\left( j,m;r\right) D_{-1,m}^{j}\left( \theta ,\varphi \right)
\end{array}
\right)  \label{21}
\end{equation}
\begin{equation}
c=\frac{1}{4\pi }\sum_{j,m}\left( 2j+1\right) \left( 
\begin{array}{c}
c_{+}\left( j,m;r\right) D_{+1,m}^{j}\left( \theta ,\varphi \right) \\ 
c_{0}\left( j,m;r\right) D_{0,m}^{j}\left( \theta ,\varphi \right) \\ 
c_{\tilde{o}}\left( j,m;r\right) D_{\tilde{o},m}^{j}\left( \theta ,\varphi
\right) \\ 
c_{-}\left( j,m;r\right) D_{-1,m}^{j}\left( \theta ,\varphi \right)
\end{array}
\right)
\end{equation}
\begin{equation}
d=\frac{1}{4\pi }\sum_{j,m}\left( 2j+1\right) \left( 
\begin{array}{c}
d_{+}\left( j,m;r\right) D_{+1,m}^{j}\left( \theta ,\varphi \right) \\ 
d_{0}\left( j,m;r\right) D_{0,m}^{j}\left( \theta ,\varphi \right) \\ 
d_{\tilde{o}}\left( j,m;r\right) D_{\tilde{o},m}^{j}\left( \theta ,\varphi
\right) \\ 
d_{-}\left( j,m;r\right) D_{-1,m}^{j}\left( \theta ,\varphi \right)
\end{array}
\right)
\end{equation}
where $D_{\lambda ,m}^{j}$ functions are the transformation matrix elements
of a general rotation. $\lambda =+1,0,\tilde{o},-1$ indices of D operator
are indicating the spin orientations of particle system.

The orthogonality relation between these operators is given as \cite{8} 
\begin{equation}
\int_{0}^{2\pi }\int_{0}^{\pi }\sin \theta d\theta D_{\lambda
,m^{`}}^{j^{\prime }\dagger }\left( \theta ,\varphi \right) D_{\lambda
,m}^{j\dagger }\left( \theta ,\varphi \right) d\theta d\varphi =\frac{4\pi }{%
2j^{`}+1}\delta _{jj^{`}}\delta _{mm^{`}}  \label{24}
\end{equation}
and the effect of helicity raising and helicity lowering operators on $D$
operator: 
\begin{eqnarray}
\partial _{-}D_{\lambda ,m}^{j} &=&\sqrt{j\left( j+1\right) }D_{\lambda
+1,m}^{j}  \nonumber \\
\partial +D_{\lambda ,m}^{j} &=&\sqrt{j\left( j+1\right) }D_{\lambda
-1,m}^{j}
\end{eqnarray}
Next by calculating both sides separately in Lagrangian it is seen that 
\begin{equation}
\sigma _{k}^{+}\partial _{-}D=\sqrt{j\left( j+1\right) }D\text{ }\sigma
_{k}^{+}
\end{equation}
and 
\begin{equation}
\sigma _{k}^{-}\partial _{+}D=\sqrt{j\left( j+1\right) }D\text{ }\sigma
_{k}^{-}  \label{27}
\end{equation}
for $k=1,2.$

For Lagrangian density the action is given in the form: 
\begin{equation}
A=\int dt\text{ }r^{2}d\widetilde{r}\sum_{jj^{`}mm^{`}}\Psi ^{\dagger
}\left( r,t\right) D_{\lambda ,m^{`}}^{j^{\prime }\dagger }\left( \widetilde{%
r}\right) [i\frac{\partial }{\partial t}-H]\text{ }D_{\lambda ,m}^{j\dagger
}\left( \widetilde{r}\right) \Psi \left( r,t\right)
\end{equation}
where 
\begin{equation}
d\widetilde{r}=\sin \theta d\theta d\varphi .
\end{equation}
We can carry out all the angular integrations with the aid of Eq.$\left( \ref
{24}\right) ,$ thereby obtaining an effective radial Lagrangian: 
\[
L\left( \overrightarrow{r}\right) =4\pi ^{2}\sum_{j,m}\left( 2j+1\right) 
\text{ }r\text{ }[a^{\dagger },b^{\dagger },c^{\dagger },d^{\dagger }]\left( 
\begin{array}{cccc}
(2\pi _{0}-m) & T_{2} & T_{1} & 0 \\ 
-T_{2} & m & 0 & -T_{1} \\ 
-T_{1} & 0 & m & -T_{2} \\ 
0 & T_{1} & T_{2} & -(2\pi _{0}+m)
\end{array}
\right) 
\]
\begin{equation}
\times r\text{ }\left( 
\begin{array}{c}
a \\ 
b \\ 
c \\ 
d
\end{array}
\right)  \label{30}
\end{equation}
where 
\begin{equation}
T_{1}=\sigma _{1}^{3}\partial _{r}-\frac{i\sqrt{j\left( j+1\right) }}{r}%
\sigma _{1}^{2}-\frac{i}{2r}\left( \overrightarrow{\sigma }_{1}\times 
\overrightarrow{\sigma }_{2}\right) ^{3},
\end{equation}
\begin{equation}
T_{2}=\sigma _{2}^{3}\partial _{r}-\frac{i\sqrt{j\left( j+1\right) }}{r}%
\sigma _{2}^{2}+\frac{i}{2r}\left( \overrightarrow{\sigma }_{1}\times 
\overrightarrow{\sigma }_{2}\right) ^{3}.  \label{32}
\end{equation}
By using the internal transformation matrices which are given as follows

\begin{equation}
\pi =\frac{1}{\sqrt{2}}\left( 
\begin{array}{cccc}
1 & 0 & 0 & 1 \\ 
0 & 1 & 1 & 0 \\ 
0 & -1 & 1 & 0 \\ 
-1 & 0 & 0 & 1
\end{array}
\right) ,\text{ \ \ \ \ }\pi ^{\dagger }=\frac{1}{\sqrt{2}}\left( 
\begin{array}{cccc}
1 & 0 & 0 & -1 \\ 
0 & 1 & -1 & 0 \\ 
0 & +1 & 1 & 0 \\ 
+1 & 0 & 0 & 1
\end{array}
\right)
\end{equation}
we define new quantities in terms of linear combinations of the original
spinor components: 
\begin{equation}
\pi \left( 
\begin{array}{c}
a \\ 
b \\ 
c \\ 
d
\end{array}
\text{ }\right) r=\frac{r}{\sqrt{2}}\left( 
\begin{array}{c}
a+d \\ 
b+c \\ 
c-b \\ 
d-a
\end{array}
\right) =\left( 
\begin{array}{c}
\Psi _{A} \\ 
\Psi _{B} \\ 
\Psi _{C} \\ 
\Psi _{D}
\end{array}
\right)  \label{34}
\end{equation}
\begin{eqnarray}
r\left( 
\begin{array}{cccc}
a^{\dagger } & b^{\dagger } & c^{\dagger } & d^{\dagger }
\end{array}
\right) \text{ }\pi ^{\dagger } &=&\frac{r}{\sqrt{2}}\left( 
\begin{array}{cccc}
\left( a^{\dagger }+d^{\dagger }\right) & \left( b^{\dagger }+c^{\dagger
}\right) & \left( c^{\dagger }-b^{\dagger }\right) & \left( d^{\dagger
}-a^{\dagger }\right)
\end{array}
\right)  \nonumber \\
&=&\left( 
\begin{array}{cccc}
\Psi _{A}^{\dagger } & \Psi _{B}^{\dagger } & \Psi _{C}^{\dagger } & \Psi
_{D}^{\dagger }
\end{array}
\right)  \label{35}
\end{eqnarray}
Since, $\pi \pi ^{\dagger }=\pi ^{\dagger }\pi =$I \ we can rewrite the
Lagrangian in the form 
\begin{equation}
L\left( \overrightarrow{r}\right) =4\pi ^{2}\sum_{j,m}\left( 2j+1\right) 
\text{ }r\text{ }[a^{\dagger },b^{\dagger },c^{\dagger },d^{\dagger }]\text{
\ }\pi ^{\dagger }\pi \left( \text{\ } 
\begin{array}{cccc}
(2\pi _{0}-m) & T_{2} & T_{1} & 0 \\ 
-T_{2} & m & 0 & -T_{1} \\ 
-T_{1} & 0 & m & -T_{2} \\ 
0 & T_{1} & T_{2} & -(2\pi _{0}+m)
\end{array}
\right) \times \pi ^{\dagger }\pi r\text{ }\left( 
\begin{array}{c}
a \\ 
b \\ 
c \\ 
d
\end{array}
\right)  \label{36}
\end{equation}
Then, with the aid of identities given in Eqs.$\left( \ref{34}\right) $, $%
\left( \ref{35}\right) $ we obtain the radial Lagrangian in the form: 
\begin{eqnarray}
L\left( \overrightarrow{r}\right) &=&4\pi ^{2}\sum_{j,m}\left( 2j+1\right)
\{\Psi _{A}^{\dagger }[-m\Psi _{A}+\left( T_{1}+T_{2}\right) \Psi _{B}-2\pi
_{0}\Psi _{D}]+\Psi _{B}^{\dagger }[m\Psi _{B}-\left( T_{1}-T_{2}\right)
\Psi _{A}]  \nonumber \\
&&+\Psi _{C}^{\dagger }[m\Psi _{C}+\left( T_{1}-T_{2}\right) \Psi _{D}]+\Psi
_{D}^{\dagger }[-2\pi _{0}\Psi _{A}+\left( T_{2}-T_{1}\right) \Psi
_{C}-m\Psi _{D}]\}  \label{37}
\end{eqnarray}

\section{Radial Equations}

The Euler-Lagrange equations of motion are 
\begin{equation}
\frac{\partial L}{\partial q^{\dagger }}-\partial _{r}\left( \frac{\partial L%
}{\partial \left( \partial _{r}q^{\dagger }\right) }\right) =0  \label{38}
\end{equation}
where $q=\Psi _{A},\Psi _{B},\Psi _{C},\Psi _{D}$ .

These yield four sets of 4-component equations, ultimately a set of sixteen
equations: 
\begin{eqnarray}
m\Psi _{A}-\left( T_{1}+T_{2}\right) \Psi _{B}+2\pi _{0}\Psi _{D} &=&0
\label{39} \\
m\Psi _{B}-\left( T_{1}+T_{2}\right) \Psi _{A} &=&0  \label{40} \\
m\Psi _{C}+\left( T_{1}-T_{2}\right) \Psi _{D} &=&0  \label{41} \\
m\Psi _{D}+\left( T_{1}-T_{2}\right) \Psi _{C}+2\pi _{0}\Psi _{A} &=&0
\label{42}
\end{eqnarray}
The explicit forms of $\left( T_{1}+T_{2}\right) $ and $\left(
T_{1}-T_{2}\right) $ are given as 
\begin{equation}
T_{1}+T_{2}=\left( 
\begin{array}{cccc}
2\partial _{r} & -\frac{\Lambda }{r} & -\frac{\Lambda }{r} & 0 \\ 
\frac{\Lambda }{r} & 0 & 0 & -\frac{\Lambda }{r} \\ 
\frac{\Lambda }{r} & 0 & 0 & -\frac{\Lambda }{r} \\ 
0 & \frac{\Lambda }{r} & \frac{\Lambda }{r} & -2\partial _{r}
\end{array}
\right) ,\text{ \ \ }T_{1}-T_{2}=\left( 
\begin{array}{cccc}
0 & \frac{\Lambda }{r} & -\frac{\Lambda }{r} & 0 \\ 
-\frac{\Lambda }{r} & 2\partial _{r} & \frac{2}{r} & -\frac{\Lambda }{r} \\ 
\frac{\Lambda }{r} & -\frac{2}{r} & -2\partial _{r} & \frac{\Lambda }{r} \\ 
0 & \frac{\Lambda }{r} & -\frac{\Lambda }{r} & 0
\end{array}
\right)  \label{43}
\end{equation}
where $\Lambda =\sqrt{j\left( j+1\right) }.$

\bigskip

Then, by defining $\Psi _{A},\Psi _{B},\Psi _{C},\Psi _{D}$ spinors as 
\[
\Psi _{A}=\left( 
\begin{array}{c}
\Psi _{A1} \\ 
\Psi _{A2} \\ 
\Psi _{A3} \\ 
\Psi _{A4}
\end{array}
\right) ,\text{ \ }\Psi _{B}=\left( 
\begin{array}{c}
\Psi _{B1} \\ 
\Psi _{B2} \\ 
\Psi _{B3} \\ 
\Psi _{B4}
\end{array}
\right) ,\text{ }\Psi _{C}=\left( 
\begin{array}{c}
\Psi _{C1} \\ 
\Psi _{C2} \\ 
\Psi _{C3} \\ 
\Psi _{C4}
\end{array}
\right) ,\text{ }\Psi _{D}=\left( 
\begin{array}{c}
\Psi _{D1} \\ 
\Psi _{D2} \\ 
\Psi _{D3} \\ 
\Psi _{D4}
\end{array}
\right) 
\]
we obtain sixteen radial equations from Eqs.$\left( \ref{39}\right) $, $%
\left( \ref{42}\right) $: 
\begin{eqnarray}
m\Psi _{A1}-2\partial _{r}\Psi _{B1}+\frac{\Lambda }{r}\Psi _{B2}+\frac{%
\Lambda }{r}\Psi _{B3}+2\pi _{0}\Psi _{D1} &=&0 \\
m\Psi _{A2}-\frac{\Lambda }{r}\Psi _{B1}+\frac{\Lambda }{r}\Psi _{B4}+2\pi
_{0}\Psi _{D2} &=&0 \\
m\Psi _{A3}-\frac{\Lambda }{r}\Psi _{B1}+\frac{\Lambda }{r}\Psi _{B4}+2\pi
_{0}\Psi _{D3} &=&0 \\
m\Psi _{A4}-\frac{\Lambda }{r}\Psi _{B2}-\frac{\Lambda }{r}\Psi
_{B3}+2\partial _{r}\Psi _{B4}+2\pi _{0}\Psi _{D4} &=&0 \\
m\Psi _{B1}-2\partial _{r}\Psi _{A1}+\frac{\Lambda }{r}\Psi _{A2}+\frac{%
\Lambda }{r}\Psi _{A3} &=&0 \\
m\Psi _{B2}-\frac{\Lambda }{r}\Psi _{A1}+\frac{\Lambda }{r}\Psi _{A4} &=&0 \\
m\Psi _{B3}-\frac{\Lambda }{r}\Psi _{A1}+\frac{\Lambda }{r}\Psi _{A4} &=&0 \\
m\Psi _{B4}-\frac{\Lambda }{r}\Psi _{A2}-\frac{\Lambda }{r}\Psi
_{A3}+2\partial _{r}\Psi _{A4} &=&0 \\
m\Psi _{C1}+\frac{\Lambda }{r}\Psi _{D2}-\frac{\Lambda }{r}\Psi _{D3} &=&0 \\
m\Psi _{C2}-\frac{\Lambda }{r}\Psi _{D1}+2\partial _{r}\Psi _{D2}+\frac{2}{r}%
\Psi _{D3}-\frac{\Lambda }{r}\Psi _{D4} &=&0 \\
m\Psi _{C3}+\frac{\Lambda }{r}\Psi _{D1}-2\partial _{r}\Psi _{D3}-\frac{2}{r}%
\Psi _{D2}+\frac{\Lambda }{r}\Psi _{D4} &=&0 \\
m\Psi _{C4}+\frac{\Lambda }{r}\Psi _{D2}-\frac{\Lambda }{r}\Psi _{D3} &=&0 \\
m\Psi _{D1}+\frac{\Lambda }{r}\Psi _{C2}-\frac{\Lambda }{r}\Psi _{C3}+2\pi
_{0}\Psi _{A1} &=&0 \\
m\Psi _{D2}-\frac{\Lambda }{r}\Psi _{C1}+2\partial _{r}\Psi _{C2}+\frac{2}{r}%
\Psi _{C3}-\frac{\Lambda }{r}\Psi _{C4}+2\pi _{0}\Psi _{A2} &=&0 \\
m\Psi _{D3}+\frac{\Lambda }{r}\Psi _{C1}-2\partial _{r}\Psi _{C3}-\frac{2}{r}%
\Psi _{C2}+\frac{\Lambda }{r}\Psi _{C4}+2\pi _{0}\Psi _{A3} &=&0 \\
m\Psi _{D4}+\frac{\Lambda }{r}\Psi _{C2}-\frac{\Lambda }{r}\Psi _{C3}+2\pi
_{0}\Psi _{A4} &=&0
\end{eqnarray}
In these equations to eliminate the equations that correspond to spin-0 we
must analysis the spin orientations of spinor wavefunctions. The spinor
wavefunction of the system must be written in terms of upper and lower
components for particle system: 
\begin{equation}
\Psi =\left( 
\begin{array}{c}
L_{1} \\ 
S_{1}
\end{array}
\right) \otimes \left( 
\begin{array}{c}
L_{2} \\ 
S_{2}
\end{array}
\right) =\left( 
\begin{array}{c}
L_{1}L_{2} \\ 
L_{1}S_{2} \\ 
S_{1}L_{2} \\ 
S_{1}S_{2}
\end{array}
\right) =\left( 
\begin{array}{c}
a \\ 
b \\ 
c \\ 
d
\end{array}
\right)  \label{60}
\end{equation}
where $a,b,c,d$ spinors are given in the form 
\begin{eqnarray}
a &=&\left( 
\begin{array}{c}
L_{1_{+}} \\ 
L_{1_{-}}
\end{array}
\right) \otimes \left( 
\begin{array}{c}
L_{2_{+}} \\ 
L_{2_{-}}
\end{array}
\right) =\left( 
\begin{array}{c}
L_{1_{+}}L_{2_{+}} \\ 
L_{1_{+}}L_{2_{-}} \\ 
L_{1_{-}}L_{2_{+}} \\ 
L_{1_{-}}L_{2_{-}}
\end{array}
\right) =\left( 
\begin{array}{c}
a_{+} \\ 
a_{0} \\ 
a_{\tilde{o}} \\ 
a_{-}
\end{array}
\right)  \label{61} \\
b &=&\left( 
\begin{array}{c}
L_{1_{+}} \\ 
L_{1_{-}}
\end{array}
\right) \otimes \left( 
\begin{array}{c}
S_{2_{+}} \\ 
S_{2_{-}}
\end{array}
\right) =\left( 
\begin{array}{c}
L_{1_{+}}S_{2_{+}} \\ 
L_{1_{+}}S_{2_{-}} \\ 
L_{1_{-}}S_{2_{+}} \\ 
L_{1_{-}}S_{2_{-}}
\end{array}
\right) =\left( 
\begin{array}{c}
b_{+} \\ 
b_{0} \\ 
b_{\tilde{o}} \\ 
b_{-}
\end{array}
\right)  \label{62} \\
c &=&\left( 
\begin{array}{c}
S_{1_{+}} \\ 
S_{1_{-}}
\end{array}
\right) \otimes \left( 
\begin{array}{c}
L_{2_{+}} \\ 
L_{2_{-}}
\end{array}
\right) =\left( 
\begin{array}{c}
S_{1_{+}}L_{2_{+}} \\ 
S_{1_{+}}L_{2_{-}} \\ 
S_{1_{-}}L_{2_{+}} \\ 
S_{1_{-}}L_{2_{-}}
\end{array}
\right) =\left( 
\begin{array}{c}
c_{+} \\ 
c_{0} \\ 
c_{\tilde{o}} \\ 
c_{-}
\end{array}
\right)  \label{63} \\
d &=&\left( 
\begin{array}{c}
S_{2_{+}} \\ 
S_{2_{-}}
\end{array}
\right) \otimes \left( 
\begin{array}{c}
S_{2_{+}} \\ 
S_{2_{-}}
\end{array}
\right) =\left( 
\begin{array}{c}
S_{1_{+}}S_{2_{+}} \\ 
S_{1_{+}}S_{2_{-}} \\ 
S_{1_{-}}S_{2_{+}} \\ 
S_{1_{-}}S_{2_{-}}
\end{array}
\right) =\left( 
\begin{array}{c}
d_{+} \\ 
d_{0} \\ 
d_{\tilde{o}} \\ 
d_{-}
\end{array}
\right)  \label{64}
\end{eqnarray}
Since our system is a single-particle system we can omit indices 1 and 2.
Hence, from Eqs.$\left( \ref{61}\right) ,\left( \ref{64}\right) $ it is seen
that 
\begin{equation}
a_{0}=a_{\tilde{o}}\text{ \ \ , \ \ }d_{0}=d_{\tilde{o}}
\end{equation}
and from Eqs.$\left( \ref{62}\right) ,\left( \ref{63}\right) $ 
\begin{equation}
b_{0}=b_{\tilde{o}}\text{ \ \ \ ,\ \ \ \ }b_{+}=c_{+}\text{ \ \ , \ }%
c_{0}=c_{\tilde{o}}\text{ \ \ \ ,\ \ \ \ \ }b_{-}=c_{-}\text{ .}
\end{equation}
Then, the new forms of $\Psi _{A},\Psi _{B},\Psi _{C},\Psi _{D}$ spinors are 
\begin{eqnarray}
\Psi _{A} &=&\frac{r}{\sqrt{2}}\left( 
\begin{array}{c}
a_{+}+d_{+} \\ 
a_{0}+d_{0} \\ 
a_{\tilde{o}}+d_{\tilde{o}} \\ 
a_{-}+d_{-}
\end{array}
\right) =\left( 
\begin{array}{c}
\Psi _{A1} \\ 
\Psi _{A2} \\ 
\Psi _{A3} \\ 
\Psi _{A4}
\end{array}
\right) \text{ \ \ \ \ \ },\text{\ \ \ \ \ }\Psi _{B}=\frac{r}{\sqrt{2}}%
\left( 
\begin{array}{c}
b_{+}+c_{+} \\ 
b_{0}+c_{0} \\ 
b_{\tilde{o}}+c_{\tilde{o}} \\ 
b_{-}+c_{-}
\end{array}
\right) =\left( 
\begin{array}{c}
\Psi _{B1} \\ 
\Psi _{B2} \\ 
\Psi _{B3} \\ 
\Psi _{B4}
\end{array}
\right) \text{\ \ } \\
\Psi _{C} &=&\frac{r}{\sqrt{2}}\left( 
\begin{array}{c}
c_{+}-b_{+} \\ 
c_{0}-b_{0} \\ 
c_{\tilde{o}}-b_{\tilde{o}} \\ 
c_{-}-b_{-}
\end{array}
\right) =\left( 
\begin{array}{c}
\Psi _{C1} \\ 
\Psi _{C2} \\ 
\Psi _{C3} \\ 
\Psi _{C4}
\end{array}
\right) \text{\ \ \ \ \ \ \ },\text{\ \ \ \ \ }\Psi _{D}=\frac{r}{\sqrt{2}}%
\left( 
\begin{array}{c}
d_{+}-a_{+} \\ 
d_{0}-a_{0} \\ 
d_{\tilde{o}}-a_{\tilde{o}} \\ 
d_{-}-a_{-}
\end{array}
\right) =\left( 
\begin{array}{c}
\Psi _{D1} \\ 
\Psi _{D2} \\ 
\Psi _{D3} \\ 
\Psi _{D4}
\end{array}
\right) \text{\ \ \ \ \ \ \ \ \ \ \ \ \ \ \ \ \ \ \ \ }
\end{eqnarray}
We can see the following equalities: 
\begin{eqnarray}
\Psi _{A2} &=&\Psi _{A3}\text{ \ , \ }\Psi _{B2}=\Psi _{B3}  \nonumber \\
\Psi _{D2} &=&\Psi _{D3}\text{ \ , \ }\Psi _{C2}=-\Psi _{C3} \\
\Psi _{C1} &=&\Psi _{C4}=0  \nonumber
\end{eqnarray}
When we use these equalities in Eqs. $\left[ 44-59\right] $ we find that
some of the solutions are identical. By eliminating one of the identitical
equations, we obtain 10 equations for spin-1 particle:

\begin{eqnarray}
m\Psi _{A1}-2\partial _{r}\Psi _{B1}+2\frac{\Lambda }{r}\Psi _{B2}+2\pi
_{0}\Psi _{D1} &=&0 \\
m\Psi _{A2}-\frac{\Lambda }{r}\Psi _{B1}+\frac{\Lambda }{r}\Psi _{B4}+2\pi
_{0}\Psi _{D2} &=&0 \\
m\Psi _{A4}-2\frac{\Lambda }{r}\Psi _{B2}+2\partial _{r}\Psi _{B4}+2\pi
_{0}\Psi _{D4} &=&0 \\
m\Psi _{B1}-2\partial _{r}\Psi _{A1}+2\frac{\Lambda }{r}\Psi _{A2} &=&0 \\
m\Psi _{B2}-\frac{\Lambda }{r}\Psi _{A1}+\frac{\Lambda }{r}\Psi _{A4} &=&0 \\
m\Psi _{B4}-2\frac{\Lambda }{r}\Psi _{A2}+2\partial _{r}\Psi _{A4} &=&0 \\
m\Psi _{C2}-\frac{\Lambda }{r}\Psi _{D1}+2\partial _{r}\Psi _{D2}+\frac{2}{r}%
\Psi _{D2}-\frac{\Lambda }{r}\Psi _{D4} &=&0 \\
m\Psi _{D1}+2\frac{\Lambda }{r}\Psi _{C2}+2\pi _{0}\Psi _{A1} &=&0 \\
m\Psi _{D2}+2\partial _{r}\Psi _{C2}-\frac{2}{r}\Psi _{C2}+2\pi _{0}\Psi
_{A2} &=&0 \\
m\Psi _{D4}+2\frac{\Lambda }{r}\Psi _{C2}+2\pi _{0}\Psi _{A4} &=&0
\end{eqnarray}

\section{Analysis of The Radial Equations}

\subsection{Asymptotic Solutions}

If we write the radial equations of spin-1 particle in the limit $%
r\rightarrow \infty $ , we obtain the following equations: 
\begin{eqnarray}
m\Psi _{A1}-2\partial _{r}\Psi _{B1}++2\pi _{0}\Psi _{D1} &=&0 \\
m\Psi _{A2}+2\pi _{0}\Psi _{D2} &=&0 \\
m\Psi _{A4}+2\partial _{r}\Psi _{B4}+2\pi _{0}\Psi _{D4} &=&0 \\
m\Psi _{B1}-2\partial _{r}\Psi _{A1} &=&0 \\
m\Psi _{B2} &=&0 \\
m\Psi _{B4}+2\partial _{r}\Psi _{A4} &=&0 \\
m\Psi _{C2}+2\partial _{r}\Psi _{D2} &=&0 \\
m\Psi _{D1}++2\pi _{0}\Psi _{A1} &=&0 \\
m\Psi _{D2}+2\partial _{r}\Psi _{C2}+2\pi _{0}\Psi _{A2} &=&0 \\
m\Psi _{D4}++2\pi _{0}\Psi _{A4} &=&0
\end{eqnarray}
The solutions of these equations are as follows: 
\begin{eqnarray}
\Psi _{A1} &=&A_{1}e^{ikr}+B_{1}e^{-ikr} \\
\Psi _{A2} &=&A_{2}e^{ikr}+B_{2}e^{-ikr} \\
\Psi _{D2} &=&A_{3}e^{ikr}+B_{3}e^{-ikr} \\
\Psi _{A2} &=&A_{4}e^{ikr}+B_{4}e^{-ikr} \\
\Psi _{D1} &=&A_{5}e^{ikr}+B_{5}e^{-ikr} \\
\Psi _{D4} &=&A_{6}e^{ikr}+B_{6}e^{-ikr} \\
\Psi _{C2} &=&A_{7}e^{ikr}+B_{7}e^{-ikr} \\
\Psi _{B1} &=&A_{8}e^{ikr}+B_{8}e^{-ikr} \\
\Psi _{B4} &=&A_{9}e^{ikr}+B_{9}e^{-ikr}
\end{eqnarray}
where 
\begin{equation}
k^{2}=E^{2}-\frac{m^{2}}{4}
\end{equation}

\subsection{Exact Solutions}

For free case the exact solutions of spin-1 particle are found in the form

\begin{eqnarray}
\Psi _{A1}\left( \rho \right) &=&\frac{1}{2}\left[ A\rho J_{j}\left( \rho
\right) +A_{1}e^{i\rho }+B_{1}e^{-i\rho }\right] \\
\Psi _{A4}\left( \rho \right) &=&\frac{1}{2}\left[ -A\rho J_{j}\left( \rho
\right) +A_{1}e^{i\rho }+B_{1}e^{-i\rho }\right] \\
\Psi _{B1}\left( \rho \right) &=&\frac{1}{2}\{D[J_{j}\left( \rho \right) +%
\frac{1}{2}\rho \left( J_{j-1}\left( \rho \right) -J_{j+1}\left( \rho
\right) \right) ]+Ee^{i\rho }+Fe^{-i\rho }\} \\
\Psi _{B4}\left( \rho \right) &=&\frac{1}{2}\{-D[J_{j}\left( \rho \right) +%
\frac{1}{2}\rho \left( J_{j-1}\left( \rho \right) -J_{j+1}\left( \rho
\right) \right) ]+Ee^{i\rho }+Fe^{-i\rho }\} \\
\Psi _{D1}\left( \rho \right) &=&\frac{1}{2}\left[ C\rho J_{j}\left( \rho
\right) +Ge^{i\rho }+He^{-i\rho }\right] \\
\Psi _{D4}\left( \rho \right) &=&\frac{1}{2}\left[ -C\rho J_{j}\left( \rho
\right) +Ge^{i\rho }+He^{-i\rho }\right] \\
\Psi _{B2} &=&\frac{\Lambda k}{\rho m}\left[ \Psi _{A1}\left( \rho \right)
-\Psi _{A4}\left( \rho \right) \right]
\end{eqnarray}
where $\rho =kr$.

\subsection{Coulomb Field Solutions}

For Coulomb field 
\begin{equation}
\pi _{0}=P_{0}-e\Phi =E+\frac{Ze^{2}}{r}.
\end{equation}
If \i t is used in spin-1 equations \ we obtain the following expreses for $%
\Psi _{A},\Psi _{B},\Psi _{C},\Psi _{D}$ spinors: 
\begin{eqnarray}
\Psi _{A1}\left( \rho \right) &=&\frac{1}{2}\left[ AR_{n,l}\left( \rho
\right) +CR_{n,l^{`}}\left( \rho \right) \right] \\
\Psi _{A4}\left( \rho \right) &=&\frac{1}{2}\left[ -AR_{n,l}\left( \rho
\right) +CR_{n,l^{`}}\left( \rho \right) \right] \\
\Psi _{D1}\left( \rho \right) &=&-\frac{1}{m}\left( E+\frac{k\alpha }{\rho }%
\right) \left[ AR_{n,l}\left( \rho \right) +CR_{n,l^{`}}\left( \rho \right) %
\right] \\
\Psi _{D4}\left( \rho \right) &=&\frac{1}{m}\left( E+\frac{k\alpha }{\rho }%
\right) \left[ AR_{n,l}\left( \rho \right) -CR_{n,l^{`}}\left( \rho \right) %
\right] \\
\Psi _{B1}\left( \rho \right) &=&\frac{k}{m}\partial _{\rho }\left[
AR_{n,l}\left( \rho \right) +CR_{n,l^{`}}\left( \rho \right) \right] \\
\Psi _{B4}\left( \rho \right) &=&\frac{k}{m}\partial _{\rho }\left[
AR_{n,l}\left( \rho \right) -CR_{n,l^{`}}\left( \rho \right) \right] \\
\Psi _{B2} &=&\frac{1}{\rho }BR_{n,l}\left( \rho \right)
\end{eqnarray}
where 
\begin{equation}
\alpha =e^{2}\text{ , \ }n=\frac{2\alpha }{k}E
\end{equation}
and 
\begin{equation}
R_{n,l}\left( \rho \right) =e^{\pm i\rho }\left[ \pm 2i\rho \right] ^{\frac{1%
}{2}\left( 1+\sqrt{\frac{l\left( l+\right) }{4}}\right) \text{ \ \ }%
}\;_{1}F_{1}\left( \frac{1}{2}+\frac{1}{4}\sqrt{l\left( l+1\right) }\mp 
\frac{in}{2},1+\frac{1}{2}\sqrt{l\left( l+1\right) },\pm 2i\rho \right)
\end{equation}
is confluent hypergeometric function.

\section{Discussion}

In relativistic dynamics the exact solutions of the wave equation are very
important because of the understanding of physics that can be brought by
such solutions. These solutions are valuable tools in determining the
radiative contributions to the energy.

Exact solutions of the Kemmer equation with potential interaction are very
rare. There has been a great deal of interest in the Kemmer equation for a
constant magnetic field but, on the other hand, we first solved exactly the
Kemmer equation for relativistic Coulomb potential. Since Kemmer equation is
a two-body Dirac-like equation it includes both spin-0 and spin-1
particles.16 solutions we obtained include both spin-0 and spin-1 particles.
The identical solutions of these 16 equations represant the spin-0 particle.
By eliminating one of the identitical equations, we obtained 10 equations
for spin-1 particle.

The asymptotic solutions of spin-1 particle are found in terms of periodic
functions. These are important in physics in order to leave the components
of spinors finite. The free solutions of spin-1 particle are found in terms
of Bessel$^{\prime }$s functions. In these solutions the $\Psi _{A1}\left(
\rho \right) ,\Psi _{B1}\left( \rho \right) ,\Psi _{D1}\left( \rho \right) $
and $\Psi _{A4}\left( \rho \right) ,\Psi _{B4}\left( \rho \right) ,\Psi
_{D4}\left( \rho \right) $ can be interpreted as helicity states of the
particle and anti-particle, respectively.

For Coulomb potential the solutions are obtained in terms of confluent
hypergeometric functions: 
\[
R_{n,l}\left( \rho \right) =e^{\pm i\rho }\left[ \pm 2i\rho \right] ^{\frac{1%
}{2}\left( 1+\sqrt{\frac{l\left( l+\right) }{4}}\right) \text{ }}\text{ }%
_{1}F_{1}\left( \frac{1}{2}+\frac{1}{4}\sqrt{l\left( l+1\right) }\mp \frac{in%
}{2},1+\frac{1}{2}\sqrt{l\left( l+1\right) },\pm 2i\rho \right) 
\]
where $l\left( l+1\right) =j\left( j+1\right) -\alpha ^{2}.$

Here $_{1}F_{1}\left( a,b;\pm 2i\rho \right) $ denotes the degenerate
hypergeometric function. For large $\rho $ this functions behaves as 
\begin{equation}
\lbrack \frac{\Gamma \left( b\right) }{\Gamma \left( a\right) }]\rho
^{a-b}e^{\rho }
\end{equation}
Demanding that F be normalizable, that is $\int_{0}^{\infty }dr$ $\left|
F\right| ^{2}=1$ implies that $\left[ \Gamma \left( a\right) \right] ^{-1}$
must vanish. This is the desired quantization condition \cite{9}. For our
solution the quantization condition is 
\begin{equation}
\frac{1}{2}\left( 1+\sqrt{\frac{j\left( j+1\right) -\alpha ^{2}}{4}}\right)
\mp \frac{i\alpha E}{\sqrt{E^{2}-\frac{m^{2}}{4}}}=-n
\end{equation}
which leads to 
\begin{equation}
E=\frac{\pm \left( m/2\right) }{\left[ 1+\ \frac{4\alpha ^{2}}{\left[ n+%
\frac{1}{2}(1+\frac{1}{2}\sqrt{\left( j+\frac{1}{2}\right) ^{2}-\left(
\alpha ^{2}+1/4\right) }\right] ^{2}}\right] ^{\frac{1}{2}}}
\end{equation}
Hence, by using the Coulomb field solutions of Kemmer equation we can
predict the energy levels of spin-1 particles. This spectra is similar to
the energy spectrum\ of spin-$\frac{1}{2}$ particle moving in Coulomb
potential which is given by 
\begin{equation}
E=\frac{\pm m}{\left[ 1+\ \frac{\alpha ^{2}}{\left[ n-\left( j+\frac{1}{2}%
\right) +\sqrt{\left( j+\frac{1}{2}\right) ^{2}-\alpha ^{2}}\right] ^{2}}%
\right] ^{\frac{1}{2}}}
\end{equation}
In a previous work\cite{10}, the bound states of a spinless charged particle
in the Coulomb field has been obtained. The radial wave functions of spin-0
particle for Coulomb interaction was obtained in terms of associated
Laguerre polynomials: 
\begin{equation}
F_{nJ}\left( \rho \right) =Ne^{-\rho /2}\rho ^{l}\text{L}_{n^{^{\prime
}}}^{2l+1}\left( \rho \right)
\end{equation}
where \ \ $n^{^{\prime }}=\lambda -l-1$ , $\lambda =n-J+l$ and N is a
normalization constant. The eigen-energy of spin-0 particle for Coulomb
potential is 
\begin{equation}
E=\frac{m}{\sqrt{1+\frac{\alpha ^{2}}{\left( n-J-l\right) ^{2}}}}
\end{equation}
From the energy spectrums of spin-1 and spin-0 particles one can see the
first order contribution $\left( j\alpha ^{2}\right) $ to the energy that
comes from the spin-Coulomb field interaction by expanding the $\left( n+%
\sqrt{\left( j+\frac{1}{2}\right) -\alpha ^{2}}\right) $ quantum number in
series in the case of spin-1. It can also be seen how these contributions
depend on spin by considering the spin-1/2 case.

The exact solutions of spin-1 particle for Coulomb potential is also
important in QED. In QED the vacuum polarization in Coulomb field is
calculated using Coulomb wave functions\cite{11}. The solutions obtained
here can be used for this purpose.

\end{document}